\documentclass[12pt,a4paper,twoside]{scrartcl}
\usepackage[pdftex]{hyperref}
\usepackage[utf8x]{inputenc}
\usepackage[utopia]{mathdesign}
\usepackage{amsmath}
\usepackage[pdftex]{graphicx}
\usepackage{titlesec}
\usepackage{tabularx}
\usepackage{textcase}
\usepackage{caption}
\usepackage{scrpage2}
\usepackage[sort&compress]{natbib}
\usepackage[all]{hypcap}

\areaset[20mm]{140mm}{220mm}
\makeatletter
\renewenvironment{table}{\@float{table}\small}{\vskip1ex\end@float}
\renewenvironment{table*}{\@dblfloat{table}\small}{\vskip1ex\end@dblfloat}
\makeatother

\newcommand{\bibfont}{\small}
\makeatletter
\newdimen\bibspacing
\renewcommand\@biblabel[1]{#1.}

\makeatother

\titleformat{\section}
{\bf}{\thesection}{1em}{\MakeTextUppercase}
\titlespacing*{\section}{0pt}{1.7\baselineskip}{0.3\baselineskip}

\titleformat{\subsection}[runin]
{\bf}{}{0em}{}[.\hspace*{0.5em}]

\newcommand   {\Ubi}        {U${}_{16}$}
\newcommand   {\Fs}         {F${}_\text{s}$}

\newcommand   {\Ca}         {C${}_{\alpha}$}

\newcommand   {\Cp}         {C${}^{\prime}$}
\newcommand   {\Cb}         {C${}_{\beta}$}
\newcommand   {\Cg}         {C${}_{\gamma}$}
\newcommand   {\Cd}         {C${}_{\delta}$}
\newcommand   {\Ce}         {C${}_{\epsilon}$}
\newcommand   {\Cz}         {C${}_{\zeta}$}
\newcommand   {\Ch}         {C${}_{\eta}$}
\newcommand   {\Sd}         {S${}_{\delta}$}
\newcommand   {\Od}         {O${}_{\delta}$}
\newcommand   {\Oe}         {O${}_{\epsilon}$}
\newcommand   {\Hz}         {H${}_{\zeta}$}
\newcommand   {\Ne}         {N${}_{\epsilon}$}
\newcommand   {\Nh}         {N${}_{\eta}$}

\newcommand   {\Eloc}       {E_\text{loc}}
\newcommand   {\Eloci}      {E_\text{loc}^\text{(1)}}
\newcommand   {\Elocii}     {E_\text{loc}^\text{(2)}}
\newcommand   {\Elociii}    {E_\text{loc}^\text{(3)}}
\newcommand   {\kloci}      {\kappa_\text{loc}^{(1)}}
\newcommand   {\klociing}   {\kappa_\text{loc}^\text{(2)}}
\newcommand   {\klociig}    {\kappa_\text{loc,G}^{(2)}}
\newcommand   {\klociii}    {\kappa_{\text{loc},i}^{(3)}}
\newcommand   {\Eev}        {E_\text{ev}}
\newcommand   {\kev}        {\kappa_\text{ev}}
\newcommand   {\Ehb}        {E_\text{hb}}
\newcommand   {\ehba}       {\epsilon^{(1)}_\text{hb}}
\newcommand   {\ehbb}       {\epsilon^{(2)}_\text{hb}}
\newcommand   {\shb}        {\sigma_\text{hb}}
\newcommand   {\Esc}        {E_\text{sc}}
\newcommand   {\Ehp}        {E_\text{hp}}
\newcommand   {\Ech}       {E_\text{ch}}

\newcommand   {\Mhp}       {M_{IJ}^{\text{(hp)}}}
\newcommand   {\Mch}       {M_{IJ}^{\text{(ch)}}}

\newcommand   {\Chp}       {C_{IJ}^{\text{(hp)}}}
\newcommand   {\Cch}       {C_{IJ}^{\text{(ch)}}}
\newcommand   {\Ahp}        {A_I^\text{(hp)}}
\newcommand   {\Ach}        {A_I^\text{(ch)}}
\newcommand   {\Ajhp}        {A_J^\text{(hp)}}
\newcommand   {\Ghp}       {\Gamma_{IJ}^\text{(hp)}}
\newcommand   {\Ghpx}       {\Gamma_{JI}^\text{(hp)}}
\newcommand   {\Tm}         {T_\text{m}}
\newcommand   {\dE}         {\Delta E}
\newcommand   {\qhb}        {q_\text{hb}}

\newcommand   {\balpha}     {$\boldsymbol{\alpha}$}
\newcommand   {\bbeta}      {$\boldsymbol{\beta}$}

\hypersetup{%
    colorlinks=true, linktocpage=true, pdfstartpage=3, pdfstartview=FitV,%
    breaklinks=true, pdfpagemode=UseNone, pageanchor=true, pdfpagemode=UseOutlines,%
    plainpages=false, bookmarksnumbered, bookmarksopen=true, bookmarksopenlevel=1,%
    hypertexnames=true, pdfhighlight=/O,
    urlcolor=black, linkcolor=black, citecolor=black,
    pdftitle={An effective all-atom potential for proteins},%
    pdfauthor={Anders Irbäck, Simon Mitternacht and Sandipan Mohanty},%
    pdfsubject={},%
    pdfkeywords={},%
    pdfcreator={pdfLaTeX},%
    pdfproducer={LaTeX with hyperref}%
}

\begin{document}
\thispagestyle{empty}
\raggedbottom
\begin{center}
  \vspace*{1cm}
  %%%%%%%%%%%%%%%%%%%%%%%%%%%%%%%%%%%%%%%%% 
  {\LARGE{\noindent An effective all-atom potential for proteins}}\\
  %%%%%%%%%%%%%%%%%%%%%%%%%%%%%%%%%%%%%%%%%% 
  \bigskip\bigskip
    Anders Irbäck\textsuperscript{1}, Simon
    Mitternacht\textsuperscript{1} and Sandipan Mohanty\textsuperscript{2}
    \bigskip\\
  {\small
    \textsuperscript{1} Computational Biology \& Biological Physics, Department of
    Theoretical Physics, Lund University, Sölvegatan 14A, SE-223 62 Lund, Sweden %
    \textsuperscript{2}J\"ulich Supercomputing Center, Institute for
    Advanced Simulation, Forschungszentrum J\"ulich, Germany\\
    \bigskip
    \textit{PMC Biophysics 2009, 2:2}
  }
\end{center}
\begin{quote}
  %\noindent%
  \small%
  We describe and test an implicit solvent all-atom potential for
  simulations of protein folding and aggregation. The potential is
  developed through studies of structural and thermodynamic properties
  of 17 peptides with diverse secondary structure. Results obtained
  using the final form of the potential are presented for all these
  peptides. The same model, with unchanged parameters, is furthermore
  applied to a heterodimeric coiled-coil system, a mixed
  $\alpha$/$\beta$ protein and a three-helix-bundle protein, with very
  good results.  The computational efficiency of the potential makes
  it possible to investigate the free-energy landscape of these
  49--67-residue systems with high statistical accuracy, using only
  modest computational resources by today's standards.
\end{quote}

\section*{Introduction}

A molecular understanding of living systems requires modeling of the
dynamics and interactions of proteins. The relevant dynamics of a
protein may amount to small fluctuations about its native structure,
or reorientations of its ordered parts relative to each other. In
either case, a tiny fraction of the conformational space is
explored. For flexible proteins, perhaps with large intrinsically
disordered parts~\cite{Uversky:02,Dyson:05}, the situation is
different.  When studying such proteins or conformational conversion
processes like folding or amyloid aggregation, the competition between
different minima on the free-energy landscape inevitably comes into
focus. Studying these systems by computer simulation is a challenge,
because proper sampling of all relevant free-energy minima must be
ensured. This goal is very hard to achieve if explicit solvent
molecules are included in the simulations.  The use of coarse-grained
models can alleviate this problem, but makes important geometric
properties like secondary structure formation more difficult to
describe.
              
Here we present an implicit solvent all-atom protein model especially
aimed at problems requiring exploration of the global free-energy
landscape.  It is based on a computationally convenient effective
potential, with parameters determined through full-scale thermodynamic
simulations of a set of experimentally well characterized
peptides. Central to the approach is the use of a single set of model
parameters, independent of the protein studied. This constraint is a
simple but efficient way to avoid unphysical biases, for example,
toward either $\alpha$-helical or $\beta$-sheet
structure~\cite{Yoda:04,Shell:08}. Imposing this constraint is also a
way to enable systematic refinement of the potential.
  
\begin{figure}
  \includegraphics[width=0.9\textwidth]{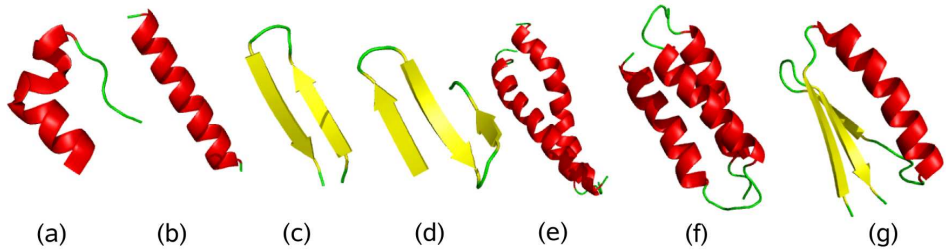}
  \caption{Schematic illustration of native geometries studied. (a) the
    Trp-cage, (b) an $\alpha$-helix, (c) a $\beta$-hairpin, (d) a
    three-stranded $\beta$-sheet, (e) an $\alpha$-helix dimer (1U2U),
    (f) a three-helix bundle (1LQ7), and (g) a mixed $\alpha/\beta$
    protein (2GJH).}\label{fig:eff_1}
\end{figure}

An earlier version~\cite{Irback:03,Irback:05a} of this potential has
proven useful, for example, for studies of
aggregation~\cite{Cheon:07,Irback:08,Li:08} and mechanical
unfolding~\cite{Irback:05b,Mitternacht:09}. Also, using a slightly
modified form of the potential~\cite{Mohanty:06}, the folding
mechanisms of a 49-residue protein, Top7-CFr, were
investigated~\cite{Mohanty:08a,Mohanty:08b}. Here we revise this
potential, through studies of an enlarged set of 17 peptides (see
Table~\ref{tab:eff_1} and Figure~\ref{fig:eff_1}). We show that the
model, in its final form, folds these different sequences to
structures similar to their experimental structures, using a single
set of potential parameters. The description of each peptide is kept
brief, to be able to discuss all systems and thereby address the issue
of transferability in a direct manner. The main purpose of this study
is model development rather than detailed characterization of
individual systems.

\begin{table}
  \caption{Amino acid sequences. Suc stands for succinylic
    acid.}\label{tab:eff_1}
  \begin{tabularx}{\textwidth}{lll}
    \hline
  System       & PDB code & Sequence \\\hline
  Trp-cage     & 1L2Y & {NLYIQ WLKDG GPSSG RPPPS} \\
  E6apn1       & 1RIJ & Ac--{ALQEL LGQWL KDGGP SSGRP PPS}--NH$_2$ \\
  C            &      & Ac--{KETAA AKFER AHA}--NH$_2$ \\
  EK           &      & Ac--{YAEAA KAAEA AKAF}--NH$_2$\\
  \Fs          &      & Suc--{AAAAA AAARA AAARA AAARA A}--NH$_2$ \\
  GCN4tp       & 2OVN & {NYHLE NEVAR LKKLV GE}   \\
  HPLC-6       & 1WFA & {DTASD AAAAA ALTAA NAKAA AELTA ANAAA AAAAT AR--NH$_2$}\\
  Chignolin    & 1UAO & {GYDPE TGTWG}         \\
  MBH12        & 1J4M & {RGKWT YNGIT YEGR}    \\
  GB1p         &      & {GEWTY DDATK TFTVT E} \\
  GB1m2        &      & {GEWTY NPATG KFTVT E} \\
  GB1m3        &      & {KKWTY NPATG KFTVQ E} \\
  trpzip1      & 1LE0 & {SWTWE GNKWT WK--NH$_2$} \\
  trpzip2      & 1LE1 & {SWTWE NGKWT WK--NH$_2$} \\
  betanova     &      & {RGWSV QNGKY TNNGK TTEGR} \\
  LLM          &      & {RGWSL QNGKY TLNGK TMEGR} \\
  beta3s       &      & {TWIQN GSTKW YQNGS TKIYT} \\
  AB zipper    & 1U2U & Ac--{EVAQL EKEVA QLEAE NYQLE QEVAQ LEHEG--NH$_2$} \\
               &      & {Ac--EVQAL KKRVQ ALKAR NYALK QKVQA LRHKG--NH$_2$} \\
  Top7-CFR     & 2GJH & {ERVRI SITAR TKKEA EKFAA ILIKV FAELG YNDIN} \\
               &      & {VTWDG DTVTV EGQL} \\
GS-$\alpha_3$W & 1LQ7 & {GSRVK ALEEK VKALE EKVKA LGGGG RIEEL KKKWE}\\
               &      & {ELKKK IEELG GGGEV KKVEE EVKKL EEEIK KL}\\
               \hline
  \end{tabularx}
\end{table}

Whether or not this potential, calibrated using data on peptides with
typically $\sim$\,20 residues, will be useful for larger systems is
not obvious. Therefore, we also apply our potential, with unchanged
parameters, to three larger systems with different geometries. These
systems are the mixed $\alpha/\beta$ protein Top7-CFr, a
three-helix-bundle protein with 67 residues, and a heterodimeric
leucine zipper composed of two 30-residue chains.
       
Protein folding simulations are by necessity based on potentials whose
terms are interdependent and dependent on the choice of geometric
representation. Therefore, we choose to calibrate our potential
directly against folding properties of whole chains. To make this
feasible, we deliberately omit many details included in force fields
like Amber, CHARMM and OPLS~(for a review, see~\cite{Ponder:03}).
With this approach, we might lose details of a given free-energy
minimum, but, by construction, we optimize the balance between
competing minima.
       
Two potentials somewhat similar in form to ours are the
$\mu$-potential of the Shakhnovich group~\cite{Hubner:05} and the PFF
potential of the Wenzel group~\cite{Herges:05}. These groups also
consider properties of entire chains for calibration, but use folded
PDB structures or sets of decoys rather than full-scale thermodynamic
simulations.  Our admittedly time-consuming procedure implies that our
model is trained on completely general structures, which might be an
advantage when studying the dynamics of folding.  Another potential
with similarities to ours is that developed by the Dokholyan group for
discrete molecular dynamics simulations~\cite{Ding:08}.

\section*{Methods} 

Our model belongs to the class of implicit solvent all-atom models
with torsional degrees of freedom. All geometrical parameters, like
bond lengths and bond angles, are as described
earlier~\cite{Irback:03}.
 
The interaction potential is composed of  four major terms: 
\begin{equation}
  E=\Eloc+\Eev+\Ehb+\Esc\,.
\label{eff:energy}
\end{equation}
The first term, $\Eloc$, contains local interactions between atoms
separated by only a few covalent bonds. The other three terms are
non-local in character: $\Eev$ represents excluded-volume effects,
$\Ehb$ is a hydrogen-bond potential, and $\Esc$ contains
residue-specific interactions between pairs of side-chains.  Next we
describe the precise form of these four terms. Energy parameters are
given in a unit called eu. The factor for conversion from eu to
kcal/mol will be determined in the next section, by calibration
against the experimental melting temperature for one of the peptides
studied, the Trp-cage.
 
\subsection*{Local potential}

The local potential $\Eloc=\Eloci+\Elocii+\Elociii$ can be divided
into two backbone terms, $\Eloci$ and $\Elocii$, and one side-chain
term, $\Elociii$. In describing the potential, the concept of a
peptide unit is useful. A peptide unit consists of the backbone \Cp O
group of one residue and the backbone NH group of the next residue.

\begin{itemize}
\item The potential $\Eloci$ represents interactions between partial
  charges of neighboring peptide units along the chain. It is given by
 \begin{equation}
   \Eloci=\kloci\sum_\textrm{n.n.}\sum_i\sum_j\frac{q_iq_j}{r_{ij}/\textrm{\AA}}\,,
   \label{eloc1}
 \end{equation} 
 where the outer sum runs over all pairs of nearest-neighbor peptide
 units and each of the two inner sums runs over atoms in one peptide
 unit (if the N side of the peptide unit is proline the sum runs over
 only \Cp\ and O). The partial charge $q_i$ is taken as $\pm$0.42 for
 \Cp\ and O atoms and $\pm$0.20 for H and N atoms. The parameter
 $\kloci$ is set to 6\,eu, corresponding to a dielectric constant of
 $\epsilon_\textrm{r}\approx$41.  Two peptide units that are not
 nearest neighbors along the chain interact through hydrogen bonding
 (see below) rather than through the potential $\Eloci$.

\item The term $\Elocii$ provides an additional OO and HH repulsion
  for neighboring peptide units, unless the residue flanked by the two
  peptide units is a glycine. This repulsion is added to make doubling
  of hydrogen bonds less likely.  Glycine has markedly different
  backbone energetics compared to other residues. The lack of
  \Cb\ atom makes glycine more flexible.  However, the observed
  distribution of Ramachandran $\phi,\psi$ angles for glycine in PDB
  structures~\cite{Hovmoller:02} is not as broad as simple steric
  considerations would suggest.  $\Elocii$ provides an energy penalty
  for glycine $\psi$ values around $\pm$120$^\circ$, which are
  sterically allowed but relatively rare in PDB structures.

  The full expression for $\Elocii$ is  
  \begin{equation}
    \Elocii=\klociing\sum_\textrm{non-Gly} \left[f(u_I)+f(v_I)\right] +
    \klociig\sum_\textrm{Gly}(\cos\psi_I+2\cos2\psi_I)\,,
    \label{eloc2}
  \end{equation}
  where $\klociing=$1.2\,eu, $\klociig=-$0.15\,eu, $I$ is a residue
  index, and
  \begin{align}
    u_I&=\min [d(\textrm{H}_I,\textrm{N}_{I+1}),d(\textrm{N}_I,\textrm{H}_{I+1})]-
    d(\textrm{H}_I,\textrm{H}_{I+1})\\
    v_I&=\min [d(\textrm{O}_I,\textrm{\Cp}_{I+1}),d(\textrm{\Cp}_I,\textrm{O}_{I+1})]-
    d(\textrm{O}_I,\textrm{O}_{I+1})\\
    f(x)&=\max(0,\tanh3x) 
  \end{align}
  
  The function $f(u_I)$ is positive if the H${}_I$H${}_{I+1}$
  distance, $d(\textrm{H}_I,\textrm{H}_{I+1})$, is smaller than both
  of the $\textrm{H}_I\textrm{N}_{I+1}$ and
  $\textrm{N}_I\textrm{H}_{I+1}$ distances, and zero otherwise.  This
  term thus provides an energy penalty when H${}_I$ and H${}_{I+1}$
  are exposed to each other (it is omitted if residue $I$ or $I+1$ is
  a proline).  Similarly, $f(v_I)$ is positive when O${}_I$ and
  O${}_{I+1}$ are exposed to each other.
 
\item $\Elociii$ is an explicit torsion angle potential for side-chain
  angles, $\chi_i$. Many side-chain angles display distributions
  resembling what one would expect based on simple steric
  considerations.  The use of the torsion potential is particularly
  relevant for $\chi_2$ in asparagine and aspartic acid and $\chi_3$
  in glutamine and glutamic acid. The torsion potential is defined as
  \begin{equation}
    \Elociii=\sum_i \klociii\cos n_i\chi_i\,,
    \label{eloc3}
  \end{equation}
  where $\klociii$ and $n_i$ are constants. Each side-chain angle
  $\chi_i$ belongs to one of four classes associated with different
  values of $\klociii$ and $n_i$ (see Table~\ref{tab:eff_2}).
\end{itemize}

\begin{table}
  \caption{Classification of side-chain angles, $\chi_i$. The
    parameters of the torsion angle potential $\Elociii$ are
    $(\klociii,n_i)=(0.6\,\textrm{eu},3)$ for class I,
    $(\klociii,n_i)=(0.3\,\textrm{eu},3)$ for class II,
    $(\klociii,n_i)=(0.4\,\textrm{eu},2)$ for class III, and
    $(\klociii,n_i)=(-0.4\,\textrm{eu},2)$ for class IV.
  }\label{tab:eff_2}
  \begin{tabularx}{\textwidth}{lcccc}
    \hline
    Residue             & $\chi_1$ & $\chi_2$ & $\chi_3$ & $\chi_4$ \\
    \hline
    Ser, Cys, Thr, Val  & I &  &  &   \\
    Ile, Leu            & I & I &  &   \\
    Asp, Asn            & I & IV &  &   \\
    His, Phe, Tyr, Trp  & I & III &  &   \\
    Met                 & I & I & II &   \\
    Glu, Gln            & I & I & IV &   \\
    Lys                 & I & I & I & I  \\
    Arg                 & I & I & I & III  \\
    \hline
  \end{tabularx}
\end{table}

\subsection*{Excluded volume}

Excluded-volume effects are modeled using the potential
\begin{equation}
  \Eev=\kev \sum_{i<j}
  \biggl[\frac{\lambda_{ij}(\sigma_i+\sigma_j)}{r_{ij}}\biggr]^{12}\,,
  \label{eff:ev}
\end{equation}
where the summation is over all pairs of atoms with a non-constant
separation, $\kev=$0.10\,eu, and $\sigma_i=$1.77, 1.75, 1.53, 1.42 and
1.00\,\AA\ for S, C, N, O and H atoms, respectively. The parameter
$\lambda_{ij}$ is unity for pairs connected by three covalent bonds
and $\lambda_{ij}=$0.75 for all other pairs. To speed up the
calculations, $\Eev$ is evaluated using a cutoff of
4.3$\lambda_{ij}$\,\AA.

\subsection*{Hydrogen bonding}

Our potential contains an explicit hydrogen-bond term, $\Ehb$.  All
hydrogen bonds in the model are between NH and CO groups. They connect
either two backbone groups or a charged side-chain (aspartic acid,
glutamic acid, lysine, arginine) with a backbone group. Two
neighboring peptide units, which interact through the local potential
(see above), are not allowed to hydrogen bond with each other.
    
The form of the hydrogen-bond potential is
\begin{equation}
  \Ehb= \ehba \sum_{{\rm bb-bb}}u(r_{ij})v(\alpha_{ij},\beta_{ij})
  +\ehbb \sum_{{\rm sc-bb}}u(r_{ij})v(\alpha_{ij},\beta_{ij})\,,
  \label{hbonds}
\end{equation}
where $\ehba=$3.0\,eu and $\ehbb=$2.3\,eu set the strengths of
backbone-backbone and sidechain-backbone bonds, respectively, $r_{ij}$
is the HO distance, $\alpha_{ij}$ is the NHO angle, and $\beta_{ij}$
is the HOC angle. The functions $u(r)$ and $v(\alpha,\beta)$ are given
by
\begin{align}
  u(r)&= 5\bigg(\frac{\shb}{r}\bigg)^{12} -
  6\bigg(\frac{\shb}{r}\bigg)^{10}\label{u}\\ v(\alpha,\beta)&=\left\{
        \begin{array}{ll}
              (\cos\alpha\cos\beta)^{1/2} &
              \ {\rm if}\ \alpha,\beta>90^{\circ}\label{v}\\
             0  & \ \mbox{otherwise}
        \end{array} \right.
\end{align}
where $\shb=2.0$\,\AA. A 4.5\,\AA\ cutoff is used for $u(r)$.

\subsection*{Side-chain potential}

Our side-chain potential is composed of two terms, $\Esc=\Ehp+\Ech$.
The $\Ech$ term represents interactions among side-chain charges. The
first and more important term, $\Ehp$, is meant to capture the effects
of all other relevant interactions, especially effective hydrophobic
attraction. For convenience, $\Ehp$ and $\Ech$ have a similar form,
\begin{equation}
  \Ehp=-\sum_{I<J}\Mhp\Chp 
  \qquad
  \Ech=-\sum_{I<J}\Mch\Cch\,.
  \label{sc2}
\end{equation}
Here the sums run over residue pairs $IJ$, $\Chp$ and $\Cch$ are
contact measures that take values between 0 and 1, and $\Mhp$ and
$\Mch$ are energy parameters.

It is assumed that ten of the twenty natural amino acids contribute to
$\Ehp$, see Table~\ref{tab:eff_3}. Included among these ten are lysine
and arginine, which are charged but have large hydrophobic parts. To
reduce the number of parameters, the hydrophobic contact energies are
taken to be additive, $\Mhp=m_I+m_J$. It is known that the
statistically derived Miyazawa-Jernigan contact
matrix~\cite{Miyazawa:96} can be approximately decomposed this
way~\cite{Li:97}. The $m_I$ parameters can be found in
Table~\ref{tab:eff_3}. $\Mhp$ is set to 0 if residues $I$ and $J$ are
nearest neighbors along the chain, and is reduced by a factor 2 for
next-nearest neighbors.

\begin{table}
  \caption{The parameter $m_I$ of the hydrophobicity potential
    $\Ehp$.}\label{tab:eff_3}
  \begin{tabularx}{\textwidth} {lc}
    \hline
    Residue & $m_I$ (eu)\\
    \hline
    Arg          		& 0.3 \\
    Met, Lys 		& 0.4 \\
    Val 			& 0.6 \\
    Ile, Leu, Pro  	        & 0.8 \\       
    Tyr                  	& 1.1 \\
    Phe, Trp        	& 1.6 \\
    \hline
  \end{tabularx}
\end{table}

The residues taken as charged are aspartic acid, glutamic acid, lysine
and arginine. The charge-charge contact energy is
$-\Mch=1.5s_Is_J$\,eu, where $s_I$ and $s_J$ are the signs of the
charges ($\pm1$).

The contact measure $\Chp$ is calculated using a predetermined set of
atoms for each amino acid, denoted by $\Ahp$ (see
Table~\ref{tab:eff_4}).  Let $n_I$ be the number of atoms in $\Ahp$
and let
\begin{equation}
  \Gamma_{IJ}^\textrm{(hp)}=\sum_{i\in \Ahp}g(\min_{j\in \Ajhp}
  r_{ij}^2)\,,
\end{equation}
where $g(x)$ is unity for $x<(3.7\textrm{\AA})^2$, vanishes for
$x>(4.5\textrm{\AA})^2$, and varies linearly for intermediate $x$.
The contact measure can then be written as 
\begin{equation}
  \Chp=\frac{\min(\gamma_{IJ}(n_I+n_J),\Ghp+\Ghpx)}{\gamma_{IJ}(n_I+n_J)}\,,
  \label{cij}
\end{equation}
where $\gamma_{IJ}$ is either 1 or 0.75.  For $\gamma_{IJ}=1$, $\Chp$
is, roughly speaking, the fraction of atoms in $\Ahp$ and $\Ajhp$ that
are in contact with some atom from the other of the two sets. A
reduction to $\gamma_{IJ}=0.75$ makes it easier to achieve a full
contact ($\Chp=1$). The value $\gamma_{IJ}=0.75$ is used for
interactions within the group proline, phenylalanine, tyrosine and
tryptophan, to make face-to-face stacking of these side-chains less
likely.  It is also used within the group isoleucine, leucine and
valine, because a full contact is otherwise hard to achieve for these
pairs.  In all other cases, $\gamma_{IJ}$ is unity.

\begin{table}
  \caption{Atoms used in the calculation of the contact measure
    $\Chp$.}\label{tab:eff_4}
  \begin{tabularx}{\textwidth} {ll}
    \hline
    Residue     & Set of atoms ($A_I$)\\
    \hline
    Pro & \Cb, \Cg, \Cd\\
    Tyr & \Cg, \Cd${}_1$, \Cd${}_2$, \Ce${}_1$, \Ce${}_2$, \Cz\\
    Val & \Cb, \Cg${}_1$, \Cg${}_2$\\
    Ile & \Cb, \Cg${}_1$, \Cg${}_2$, \Cd\\ 
    Leu & \Cb, \Cg, \Cd${}_1$, \Cd${}_2$\\
    Met & \Cb, \Cg, \Sd, \Ce\\
    Phe & \Cg, \Cd${}_1$, \Cd${}_2$, \Ce${}_1$, \Ce${}_2$, \Cz\\        
    Trp & \Cg, \Cd${}_1$, \Cd${}_2$, \Ce${}_3$, \Cz${}_3$, \Ch${}_2$\\          
    Arg & \Cb, \Cg\\
    Lys & \Cb, \Cg, \Cd\\
    \hline
  \end{tabularx}
\end{table}

The definition of $\Cch$ is similar. The $\gamma_{IJ}$ parameter is
unity for charge-charge interactions, and the sets of atoms used,
$\Ach$, can be found in Table~\ref{tab:eff_5}.

\begin{table}
  \caption{Atoms used in the calculation of the contact measure
    $\Cch$.}\label{tab:eff_5}
  \begin{tabularx}{\textwidth} {ll}
    \hline
    Residue     & Set of atoms ($A_I$)\\
    \hline
    Arg & \Ne, \Cz, \Nh${}_1$, \Nh${}_2$\\
    Lys & ${}^1$\Hz, ${}^2$\Hz, ${}^3$\Hz\\
    Asp & \Od${}_1$, \Od${}_2$\\
    Glu & \Oe${}_1$, \Oe${}_2$\\
    \hline
  \end{tabularx}
\end{table}

\subsection*{Chain ends}

Some of the sequences we study have extra groups attached at one or
both ends of the chain. The groups occurring are N-terminal acetyl and
succinylic acid, and C-terminal NH${}_2$. When such a unit is present,
the model assumes polar NH and CO groups beyond the last \Ca\ atom to
hydrogen bond like backbone NH/CO groups but with the strength reduced
by a factor 2 (multiplicatively). The charged group of succinylic acid
interacts like a charged side-chain.
        
In the absence of end groups, the model assumes the N and C termini to
be positively and negatively charged, respectively, and to interact
like charged side-chains.
 
\subsection*{Monte Carlo details}

We investigate the folding thermodynamics of this model by Monte Carlo
(MC) methods. The simulations are done using either simulated
tempering (ST)~\cite{Lyubartsev:92,Marinari:92} or parallel
tempering/replica exchange (PT)~\cite{Swendsen:86,Hukushima:96}, both
with temperature as a dynamical variable. For small systems we use ST,
with seven geometrically distributed temperatures in the range
279\,K--367\,K. For each system, ten independent ST runs are
performed. For our largest systems we use PT with a set of sixteen
temperatures, spanning the same interval. Using fourfold
multiplexing~\cite{Meinke:09a}, one run comprising 64 parallel
trajectories is performed for each system.  The PT temperature
distribution is determined by an optimization
procedure~\cite{Meinke:09a}. The length of our different simulations
can be found in Table~\ref{tab:eff_6}.

\begin{table}
  \caption{Algorithm used and total number of elementary MC steps for
    all systems studied.}\label{tab:eff_6}
  \begin{tabularx}{\textwidth}{lll}
    \hline
    System           & Method & MC steps \\
    \hline
    Trp-cage, E6apn1  & ST & $10\times1.0\times10^9$ \\
    C, EK, \Fs, GCN4tp& ST & $10\times1.0\times10^9$ \\
    HPLC-6            & ST & $10\times3.0\times10^9$ \\
    Chignolin         & ST & $10\times0.5\times10^9$ \\
    MBH12             & ST & $10\times1.0\times10^9$ \\
    GB1p              & ST & $10\times2.0\times10^9$ \\
    GB1m2, GB1m3      & ST & $10\times1.0\times10^9$ \\
    trpzip1, trpzip2  & ST & $10\times1.0\times10^9$ \\
    betanova, LLM     & ST & $10\times1.0\times10^9$ \\
    beta3s            & ST & $10\times2.0\times10^9$ \\
    AB zipper         & PT & $64\times3.0\times10^9$ \\
    Top7-CFR          & PT & $64\times2.4\times10^9$ \\
    GS-$\alpha_3$W    & PT & $64\times3.5\times10^9$  \\
    \hline
  \end{tabularx}
\end{table}

Three different conformational updates are used in the simulations:
single variable updates of side-chain and backbone angles,
respectively, and Biased Gaussian Steps (BGS)~\cite{Favrin:01}.  The
BGS move is semi-local and updates up to eight consecutive backbone
degrees of freedom in a manner that keeps the ends of the segment
approximately fixed. The ratio of side-chain to backbone updates is the
same at all temperatures, whereas the relative frequency of the two
backbone updates depends on the temperature. At high temperatures the
single variable update is the only backbone update used, and at low
temperatures only BGS is used. At intermediate temperatures both
updates are used.

The AB zipper, a two-chain system, is studied using a periodic box of
size $(158\,\textrm{\AA})^3$. In addition to the conformational
updates described above, the simulations of this system used rigid
body translations and rotations of individual chains.
   
Our simulations are performed using the open source C++-package
PROFASI~\cite{profasi}. Future public releases of PROFASI will include
an implementation of the force field described here. While this force
field has been implemented in PROFASI in an optimized manner, this
optimization does not involve a parallel evaluation of the potential
on many processors. Therefore, in our simulations the number of
processors used is the same as the number of MC trajectories
generated. For a typical small peptide, a trajectory of the length as
given in Table~\ref{tab:eff_6} takes $\sim$\,18 hours to generate on
an AMD Opteron processor with $\sim$\,2.0 GHz clock rate. For the
largest system studied, GS-$\alpha_3$W, the simulations, with a
proportionately larger number of MC updates, take $\sim$\,10 days to
complete.

\subsection*{Analysis}

In our simulations, we monitor a variety of different properties.
Three important observables are as follows.
\begin{enumerate}
\item $\alpha$-helix content, $h$. A residue is defined as helical if
  its Ramachandran angle pair is in the region
  $-90^\circ<\phi<-30^\circ$, $-77^\circ<\psi<-17^\circ$.
  Following~\cite{Garcia:02}, a stretch of $n>2$ helical residues is
  said to form a helical segment of length $n-2$. For an end residue
  that is not followed by an extra end group, the $(\phi,\psi)$ pair
  is poorly defined. Thus, for a chain with $N$ residues, the maximum
  length of a helical segment is $N-4$, $N-3$ or $N-2$, depending on
  whether there are zero, one or two end groups.  The $\alpha$-helix
  content $h$ is defined as the total length of all helical segments
  divided by this maximum length.
\item Root-mean-square deviation from a folded reference structure,
  bRMSD\slash RMSD\slash pRMSD. bRMSD is calculated over backbone
  atoms, whereas RMSD is calculated over all heavy atoms. All residues
  except the two end residues are included in the calculation, unless
  otherwise stated. For the case of the dimeric AB zipper, the
  periodic box used for the simulations has to be taken into account.
  The two chains in the simulation might superficially appear to be
  far away when they are in fact close, because of periodicity. For
  this case we evaluate backbone RMSD over atoms taken from both
  chains in the dimer, and minimize this value with respect to
  periodic translations. We denote this as pRMSD.
\item Nativeness measure based on hydrogen bonds, $\qhb$. This
  observable has the value 1 if at most two native backbone-backbone
  hydrogen bonds are missing, and is 0 otherwise. A hydrogen bond is
  considered formed if its energy is less than $-1.03$\,eu.
\end{enumerate}

In many cases, it turns out that the temperature dependence 
of our results can be approximately described in terms of the simple 
two-state model 
\begin{equation}
  X(T)=\frac{X_1+X_2 K(T)}{1+K(T)} \qquad
  K(T)=\exp\left[\left(\frac{1}{RT}-\frac{1}{R\Tm}\right)\dE\right]\
  \label{twostate}
\end{equation}
where $X(T)$ is the quantity studied, $X_1$ and $X_2$ are the values
of $X$ in the two states, and $K(T)$ is the effective equilibrium
constant ($R$ is the gas constant).  In this first-order form, $K(T)$
contains two parameters: the melting temperature $\Tm$ and the energy
difference $\dE$.  The parameters $\Tm$, $\dE$, $X_1$ and $X_2$ are
determined by fitting to data.

Thermal averages and their statistical errors are calculated by using
the jackknife method~\cite{Miller:74}, after discarding the first
20\,\% of each MC trajectory for thermalization.

Figures of 3D structures were prepared using
PyMOL~\cite{pymol}.

\section*{Results}

We study a total of 20 peptide/protein systems, listed in
Table~\ref{tab:eff_1} (amino acid sequences can be found in this
table). Among these, there are 17 smaller systems with 10--37 residues
and 3 larger ones with $\ge49$ residues.  Many of the smaller systems
have been simulated by other groups, in some cases with explicit water
(for a review, see~\cite{Gnanakaran:03}). Two of the three larger
systems, as far as we know, have not been studied using other force
fields. A study of the 67-residue three-helix-bundle protein
GS-$\alpha_3$W using the ECEPP/3 force field was recently
reported~\cite{Meinke:09b}.  The simulations presented here use the
same geometric representation and find about a hundred times the
number of independent folding events, while consuming much smaller
computing resources.

\subsection*{Trp-cage and E6apn1}

The Trp-cage is a designed 20-residue miniprotein with a compact
helical structure~\cite{Neidigh:02}. Its NMR-derived native structure
(see Figure~\ref{fig:eff_1}) contains an $\alpha$-helix and a single
turn of $3_{10}$-helix~\cite{Neidigh:02}.  The E6apn1 peptide was
designed using the Trp-cage motif as a scaffold, to inhibit the E6
protein of papillomavirus~\cite{Liu:04}. E6apn1 is three residues
larger than the Trp-cage but has a similar structure, except that the
$\alpha$-helix is slightly longer~\cite{Liu:04}.

As indicated earlier, we use melting data for the Trp-cage to set the
energy scale of the model. For this peptide, several experiments found
a similar melting temperature, $\Tm\sim
315$\,K~\cite{Neidigh:02,Qiu:02,Streicher:07}. In our model, the heat
capacity of the Trp-cage displays a maximum at $RT=0.4722 \pm 0.0008$
eu. Our energy unit eu is converted to kcal/mol by setting this
temperature equal to the experimental melting temperature (315\,K).
Having done that, there is no free parameter left in the model. Other
systems are thus studied without tuning any model parameter. For
E6apn1, the experimental melting temperature is
$\Tm\sim305$\,K~\cite{Liu:04}.

\begin{figure}
  \includegraphics[width=\textwidth]{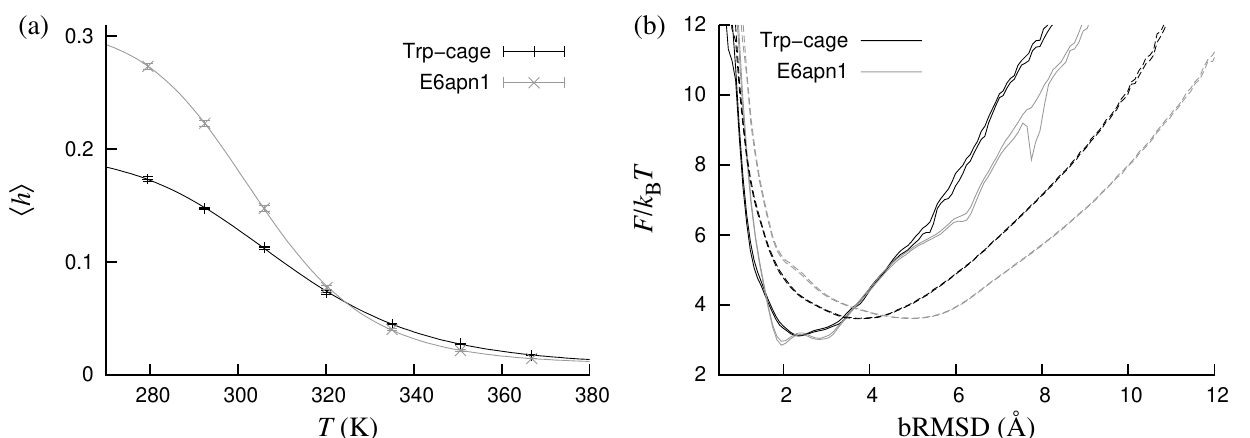}
  \caption{The Trp-cage and E6apn1. (a) Helix content $h$ against
    temperature.  The lines are two-state fits ($\Tm=309.6\pm0.7$\,K
    and $\dE=11.3\pm0.3$\,kcal/mol for the Trp-cage;
    $\Tm=304.0\pm0.5$\,K and $\dE=14.2\pm0.3$\,kcal/mol for E6apn1).
    (b) Free energy $F$ calculated as a function of bRMSD at two
    different temperatures, 279\,K (solid lines) and 306\,K (dashed
    lines). The double lines indicate the statistical
    errors.}\label{fig:eff_2}
\end{figure}

Figure~\ref{fig:eff_2}a shows the helix content $h$ against
temperature for the Trp-cage and E6apn1, as obtained from our
simulations.  In both cases, the $T$ dependence is well described by
the simple two-state model of Equation~\ref{twostate}.  The fitted
melting temperatures are $\Tm=309.6\pm0.7$\,K and $\Tm=304.0\pm0.5$\,K
for the Trp-cage and E6apn1, respectively. This $\Tm$ value for the
Trp-cage is slightly lower than that we obtain from heat capacity
data, 315\,K.  A fit to our data for the hydrophobicity energy $\Ehp$
(not shown) gives instead a slightly larger $\Tm$, $321.1\pm0.8$\,K.
This probe dependence of $\Tm$ implies an uncertainty in the
determination of the energy scale. By using the Trp-cage, this
uncertainty is kept small ($\sim$\,2\,\%). For many other peptides,
the spread in $\Tm$ is much larger (see below).

Figure~\ref{fig:eff_2}b shows the free energy calculated as a function
of bRMSD for the Trp-cage and E6apn1 at two different
temperatures. The first temperature, 279\,K, is well below $\Tm$. Here
native-like conformations dominate and the global free-energy minima
are at 2.4\,\AA\ and 2.0\,\AA\ for the Trp-cage and E6apn1,
respectively.  At the second temperature, 306\,K, the minima are
shifted to higher bRMSD. Note that these free-energy
profiles, taken near $\Tm$, show no sign of a double-well
structure. Hence, these peptides do not show a genuine two-state
behavior in our simulations, even though the melting curves
(Figure~\ref{fig:eff_2}a) are well described by a two-state model, as
are many experimentally observed melting curves.

\subsection*{The \balpha-helices C, EK, \Fs, GCN4tp and HPLC-6}

Our next five sequences form $\alpha$-helices. Among these, there are
large differences in helix stability, according to CD studies.  The
least stable are the C~\cite{Bierzynski:82} and EK~\cite{Scholtz:95}
peptides, which are only partially stable at $T\sim 273$\,K.  The
original C peptide is a 13-residue fragment of ribonuclease A, but the
C peptide here is an analogue with two alanine substitutions and a
slightly increased helix stability~\cite{Shoemaker:87}.  The EK
peptide is a designed alanine-based peptide with 14 residues.

Our third $\alpha$-helix peptide is the 21-residue
\Fs~\cite{Lockhart:92}, which is also alanine-based. \Fs\ is more
stable than C and EK~\cite{Lockhart:92,Lockhart:93}, with estimated
$\Tm$ values of 308\,K~\cite{Lockhart:93} and
303\,K~\cite{Thompson:97} from CD studies and 334\,K from an IR
study~\cite{Williams:96}.  Even more stable is HPLC-6, a winter
flounder antifreeze peptide with 37 residues.  CD data suggest that
the helix content of HPLC-6 remains non-negligible, $\sim0.10$, at
temperatures as high as $\sim$\,343\,K~\cite{Chakrabartty:89}. Our
fifth helix-forming sequence, which we call GCN4tp, has 17
residues and is taken from a study of GCN4 coiled-coil
formation~\cite{Steinmetz:07}. Its melting behavior has not been
studied, as far as we know, but its structure was characterized by
NMR~\cite{Steinmetz:07}.

\begin{figure}
  \centering
  \includegraphics[width=0.5\textwidth]{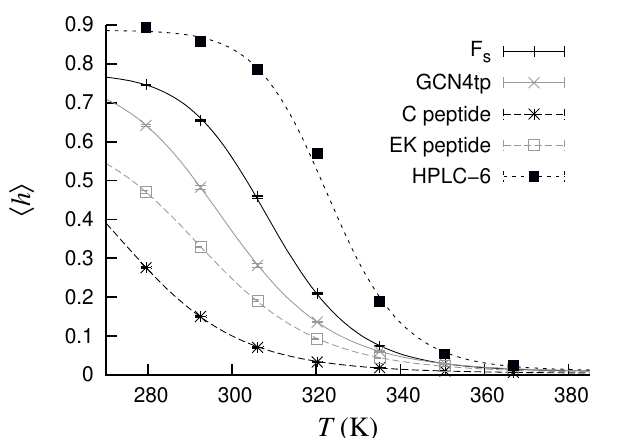}
  \caption{The C, EK, \Fs, GCN4tp and HPLC-6 peptides.  Helix content
    $h$ against temperature. The lines are two-state fits
    ($\Tm=276.3\pm2.4$\,K and $\dE=11.7\pm0.4$\,kcal/mol for C;
    $\Tm=293.9\pm0.4$\,K and $\dE=12.6\pm0.2$\,kcal/mol for
    {EK}; $\Tm=309.2\pm0.3$\,K and
    $\dE=18.7\pm0.4$\,kcal/mol for \Fs; $\Tm=298.9\pm0.1$\,K and
    $\dE=14.1\pm0.1$\,kcal/mol for {GCN4}tp;
    $\Tm=323.3\pm1.2$\,K and $\dE=23.6\pm2.2$\,kcal/mol for
    {HPLC-6}).}\label{fig:eff_3}
\end{figure}

These five peptides are indeed $\alpha$-helical in our model. At
279\,K, the calculated helix content $h$ is 0.28 for the C peptide,
0.47 for the EK peptide, and $>$\,0.60 for the other three peptides.
Figure~\ref{fig:eff_3} shows the temperature dependence of $h$. By
fitting Equation~\ref{twostate} to the data for the three stable
sequences, we find melting temperatures of $298.9\pm0.1$\,K,
$309.2\pm0.3$\,K and $323.3\pm1.2$\,K for GCN4tp, \Fs\ and HPLC-6,
respectively.

For the four peptides whose melting behavior has been studied
experimentally, these results are in good agreement with experimental
data.  In particular, we find that HPLC-6 indeed is more stable than
\Fs\ in the model, which in turn is more stable than both C and EK.
The model thus captures the stability order among these peptides.

\subsection*{The \bbeta-hairpins chignolin and MBH12}

We now turn to $\beta$-sheet peptides and begin with the
$\beta$-hairpins chignolin~\cite{Honda:04} and MBH12~\cite{Pastor:02}
with 10 and 14 residues, respectively. Both are designed and have been
characterized by NMR. For chignolin, $\Tm$ values in the range
311--315\,K were reported~\cite{Honda:04}, based on CD and NMR.  We
are not aware of any melting data for MBH12.

\begin{figure}
  \includegraphics[width=\textwidth]{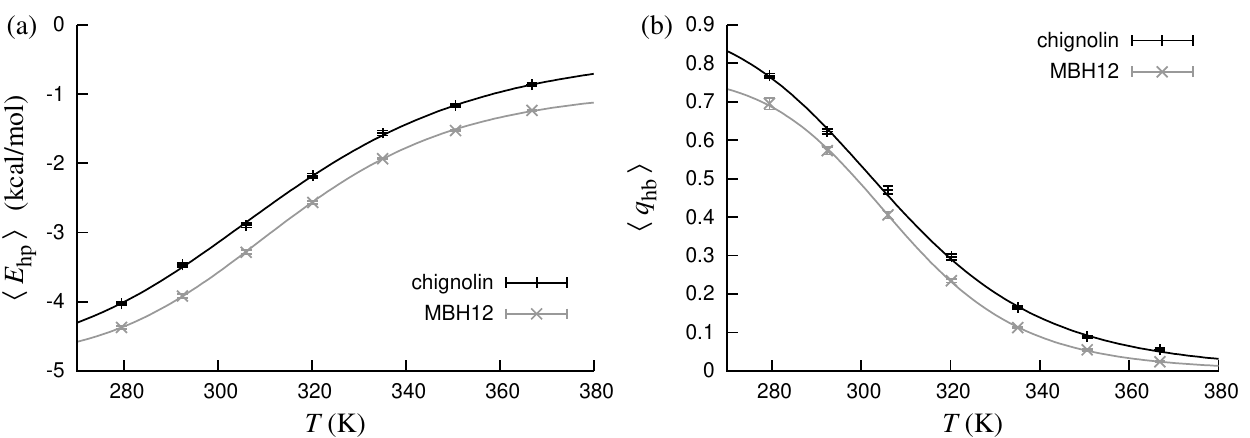}
  \caption{Chignolin and MBH12. (a) Hydrophobicity energy $\Ehp$
    against temperature. The lines are two-state fits
    ($\Tm=311.0\pm0.5$\,K and $\dE=9.6\pm0.2$\,kcal/mol for chignolin;
    $\Tm=315.4\pm1.3$\,K and $\dE=9.9\pm0.9$\,kcal/mol for
    MBH12).  (b) Nativeness $\qhb$ against temperature.
    The lines are two-state fits ($\Tm=305.4\pm0.5$\,K and
    $\dE=10.4\pm0.1$\,kcal/mol for chignolin; $\Tm=309.2\pm0.7$\,K and
    $\dE=13.5\pm0.2$\,kcal/mol for MBH12).
  }\label{fig:eff_4}
\end{figure}

Figure~\ref{fig:eff_4} shows the temperature dependence of the
hydrophobicity energy $\Ehp$ and the nativeness parameter $\qhb$ for
these peptides.  By fitting to $\Ehp$ data, we obtain
$\Tm=311.0\pm0.5$\,K and $\Tm=315.4\pm1.3$\,K for chignolin and MBH12,
respectively.  Using $\qhb$ data instead, we find $\Tm=305.4\pm0.5$\,K
for chignolin and $\Tm=309.2\pm0.7$\,K for MBH12. These $\Tm$ values
show a significant but relatively weak probe dependence. The values
for chignolin can be compared with experimental data, and the
agreement is good.

Because these peptides have only four native hydrogen bonds each, one
may question our definition of $\qhb$ (see Methods), which takes a
conformation as native-like ($\qhb=1$) even if two hydrogen bonds are
missing. Therefore, we repeated the analysis using the stricter
criterion that native-like conformations ($\qhb=1$) may lack at most
one hydrogen bond. The resulting decrease in native population, as
measured by the average $\qhb$, was $\sim$\,0.1 or smaller at all
temperatures. Even with this stricter definition, we find native
populations well above 0.5 at low temperatures for both peptides.

\subsection*{The \bbeta-hairpins GB1p, GB1m2 and GB1m3}

GB1p is the second $\beta$-hairpin of the B1 domain of
protein G (residues 41--56). Its folded population has been estimated
by CD/NMR to be 0.42 at 278\,K~\cite{Blanco:94} and $\sim$\,0.30 at
298\,K~\cite{Fesinmeyer:04}, whereas a Trp fluorescence study found a
$\Tm$ of 297\,K~\cite{Munoz:97}, corresponding to a somewhat higher
folded population. GB1m2 and GB1m3 are two
mutants of {GB1}p with significantly enhanced
stability~\cite{Fesinmeyer:04}. At 298\,K, the folded population was
found to be $0.74\pm0.05$ for {GB1}m2 and $0.86\pm0.03$ for
{GB1}m3, based on CD and NMR
measurements~\cite{Fesinmeyer:04}. It was further estimated that
$\Tm=320\pm2$\,K for {GB1}m2 and $\Tm=333\pm2$\,K for
{GB1}m3~\cite{Fesinmeyer:04}.

All these three peptides are believed to adopt a structure similar to
that {GB1}p has as part of the protein G B1 domain (PDB code
1{GB1}). This part of the full protein contains seven
backbone-backbone hydrogen bonds. These hydrogen bonds are the ones we
consider when evaluating $\qhb$ for these peptides.
 
\begin{figure}
  \includegraphics[width=\textwidth]{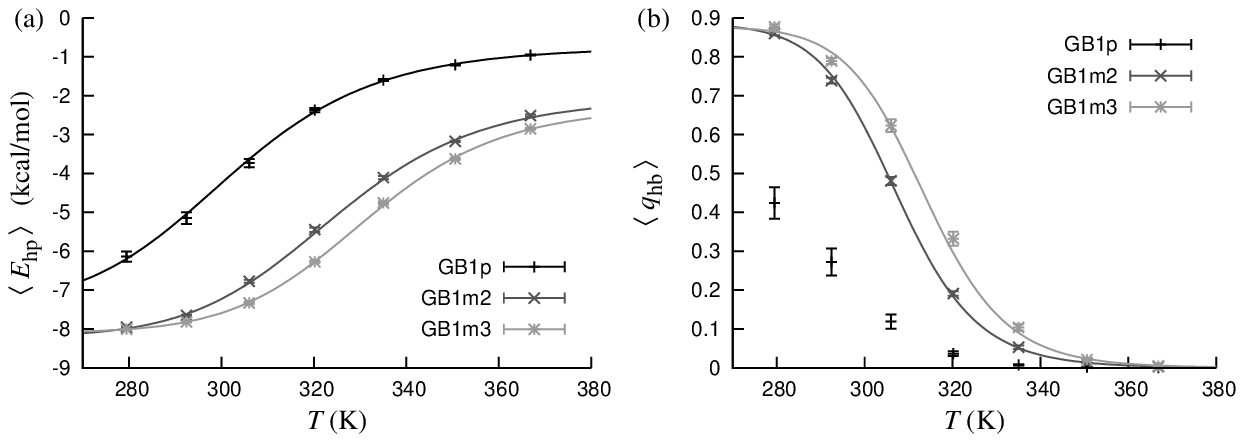}
  \caption{{GB1}p, {GB1}m2 and
    {GB1}m3. (a) Hydrophobicity energy $\Ehp$ against
    temperature.  The lines are two-state fits ($\Tm=301.7\pm3.3$\,K
    and $\dE=11.3\pm 1.1$\,kcal/mol for {GB1}p;
    $\Tm=324.4\pm1.4$\,K and $\dE=13.2\pm1.0$\,kcal/mol for
    {GB1}m2; $\Tm=331.4\pm0.7$\,K and
    $\dE=14.8\pm0.5$\,kcal/mol for {GB1}m3).  (b) Nativeness
    $\qhb$ against temperature.  The lines are two-state fits
    ($\Tm=307.5\pm0.5$\,K and $\dE=20.7\pm0.5$\,kcal/mol for
    {GB1}m2; $\Tm=313.9\pm1.4$\,K and
    $\dE=21.4\pm1.1$\,kcal/mol for {GB1}m3).
  }\label{fig:eff_5}
\end{figure}

Figure~\ref{fig:eff_5} shows the observables $\Ehp$ and $\qhb$ against
temperature for these peptides. Fits to the data give $\Ehp$-based
$\Tm$ values of $301.7\pm3.3$\,K, $324.4\pm1.1$\,K and
$331.4\pm0.7$\,K for {GB1}p, {GB1}m2 and
{GB1}m3, respectively, and $\qhb$-based $\Tm$ values of
$307.5\pm0.5$\,K and $313.9\pm1.4$\,K for {GB1}m2 and
{GB1}m3, respectively.  The $\qhb$ data do not permit a
reliable fit for the less stable {GB1}p.  At 298\,K, we find
$\qhb$-based folded populations of 0.20, 0.64 and 0.74 for
{GB1}p, {GB1}m2 and {GB1}m3,
respectively, which can be compared with the above-mentioned
experimental results (0.30, 0.74 and 0.86).

These results show that, in the model, the apparent folded populations
of these peptides depend quite strongly on the observable studied. Our
$\Ehp$-based results agree quite well with experimental data,
especially for {GB1}m2 and {GB1}m3, whereas our
$\qhb$ results consistently give lower folded populations for all
peptides. The stability order is the same independent of which of the
two observables we study, namely
{GB1}p\,$<$\,{GB1}m2$\,<\,${GB1}m3,
which is the experimentally observed order.

The stability difference between {GB1}m2 and
{GB1}m3 is mainly due to charge-charge interactions. In our
previous model~\cite{Irback:05a}, these interactions were ignored, and
both peptides had similar stabilities.  The present model splits this
degeneracy. Moreover, the magnitude of the splitting, which
sensitively depends on the strength of the charge-charge interactions,
is consistent with experimental data.

\subsection*{The \bbeta-hairpins trpzip1 and trpzip2}

The 12-residue trpzip1 and trpzip2 are designed $\beta$-hairpins, each
containing two tryptophans per $\beta$-strand~\cite{Cochran:01}. The
only difference between the two sequences is a transposition of an
asparagine and a glycine in the hairpin turn. CD measurements suggest
that trpzip1 and trpzip2 are remarkably stable for their size, with
$\Tm$ values of $323$\,K and $345$\,K, respectively~\cite{Cochran:01}.
A complementary trpzip2 study, using both experimental and
computational methods, found $\Tm$ values to be strongly
probe-dependent~\cite{Yang:04}.

\begin{figure}
  \includegraphics[width=\textwidth]{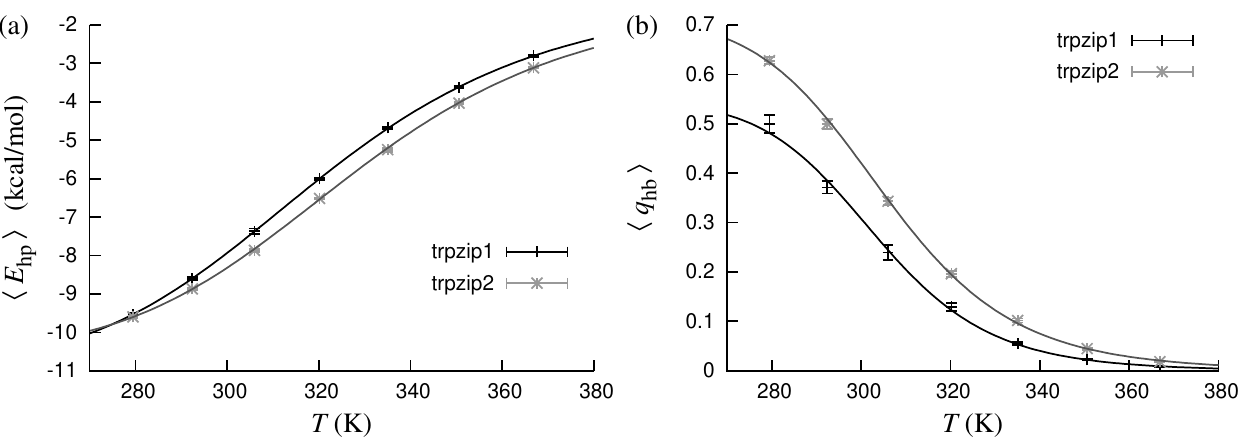}
  \caption{Trpzip1 and trpzip2. (a) Hydrophobicity energy $\Ehp$
    against temperature. The lines are two-state fits
    ($\Tm=319.7\pm0.2$\,K and $\dE=7.9\pm0.1$\,kcal/mol for trpzip1;
    $\Tm=327.1\pm0.8$\,K and $\dE=8.3\pm0.4$\,kcal/mol for trpzip2).
    (b) Nativeness $\qhb$ against temperature.  The lines are
    two-state fits ($\Tm=303.2\pm1.8$\,K and
    $\dE=14.1\pm0.5$\,kcal/mol for trpzip1; $\Tm=305.0\pm1.1$\,K and
    $\dE=12.6\pm0.3$\,kcal/mol for trpzip2).  }\label{fig:eff_6}
\end{figure}

Figure~\ref{fig:eff_6} shows our melting curves for these peptides,
based on the observables $\Ehp$ and $\qhb$. The $\Ehp$-based $\Tm$
values are $319.7\pm0.2$\,K and $327.1\pm0.8$\,K for trpzip1 and
trpzip2, respectively. Using $\qhb$ data instead, we find
$\Tm=303.2\pm1.1$\,K for trpzip1 and $\Tm=305.0\pm1.1$\,K for trpzip2.

Like for the other $\beta$-hairpins discussed earlier, our
$\qhb$-based folded populations are low compared to estimates based on
CD data, whereas those based on $\Ehp$ are much closer to
experimental data. For trpzip2, the agreement is not perfect but
acceptable, given that $\Tm$ has been found to be strongly
probe-dependent for this peptide~\cite{Yang:04}.

\subsection*{Three-stranded \bbeta-sheets: betanova, LLM and beta3s}

Betanova~\cite{Kortemme:98}, the betanova triple mutant
LLM~\cite{Lopez:01} and beta3s~\cite{deAlba:99} are designed
20-residue peptides forming three-stranded $\beta$-sheets. All the
three peptides are marginally stable.  NMR studies suggest that the
folded population at 283\,K is 0.09 for betanova~\cite{Lopez:01}, 0.36
for LLM~\cite{Lopez:01}, and 0.13--0.31 for beta3s~\cite{deAlba:99}.

\begin{figure}
  \includegraphics[width=\textwidth]{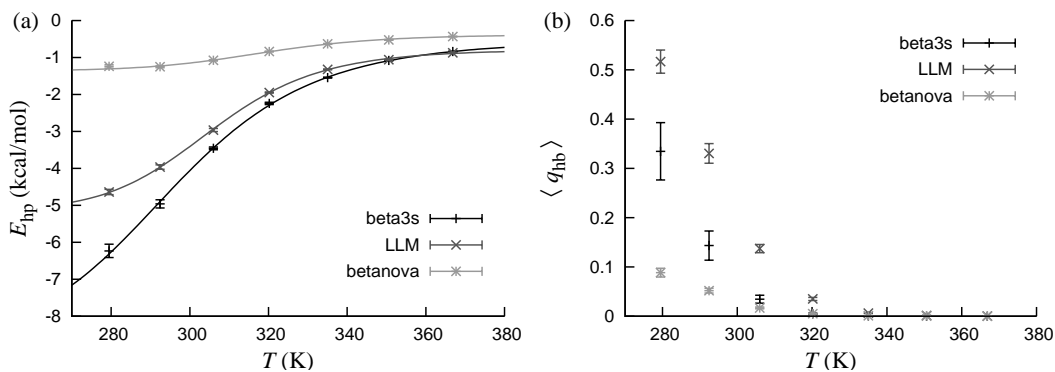}
  \caption{Betanova, LLM and beta3s.  (a) Hydrophobicity energy $\Ehp$
    against temperature.  The lines are two-state fits
    ($\Tm=318.8\pm2.5$\,K and $\dE=13.3\pm2.1$\,kcal/mol for betanova;
    $\Tm=305.6\pm1.7$\,K and $\dE=13.4\pm1.0$\,kcal/mol for LLM;
    $\Tm=295.7\pm3.1$\,K and $\dE=9.7\pm0.5$\,kcal/mol for beta3s).
    (b) Nativeness $\qhb$ against temperature.  Two-state fits were
    not possible.}\label{fig:eff_7}
\end{figure}

Figure~\ref{fig:eff_7} shows our $\Ehp$ and $\qhb$ data for these
peptides.  From the $\qhb$ data, $\Tm$ values cannot be extracted,
because the stability of the peptides is too low. At 283\,K, the
$\qhb$-based folded populations are 0.08, 0.47, 0.28 for betanova, LLM
and beta3s, respectively, in good agreement with the experimental
results. Fits to $\Ehp$ data can be performed. The obtained $\Tm$
values are $318.8\pm2.5$\,K, $305.6\pm1.7$\,K and $295.7\pm 3.1$\,K
for betanova, LLM and beta3s, respectively.

These $\Ehp$-based $\Tm$ values are high compared to the
experimentally determined folded populations, especially for
betanova. Note that betanova has a very low hydrophobicity. The
correlation between $\Ehp$ and folding status is therefore likely to
be weak for this peptide.

In contrast to the $\Ehp$-based folded populations, those based on
$\qhb$ agree quite well with experimental data. In this respect, the
situation is the opposite to what we found for the $\beta$-hairpins
studied above.  A possible reason for this difference is discussed
below.
            
\subsection*{AB zipper}

The AB zipper is a designed heterodimeric leucine zipper, composed of
an acidic A chain and a basic B chain, each with 30
residues~\cite{Marti:04}.  The dimer structure has been characterized
by NMR, and a melting temperature of $\sim$\,340\,K was estimated by
CD measurements (at neutral pH)~\cite{Marti:04}.

\begin{figure}
  \includegraphics[width=\textwidth]{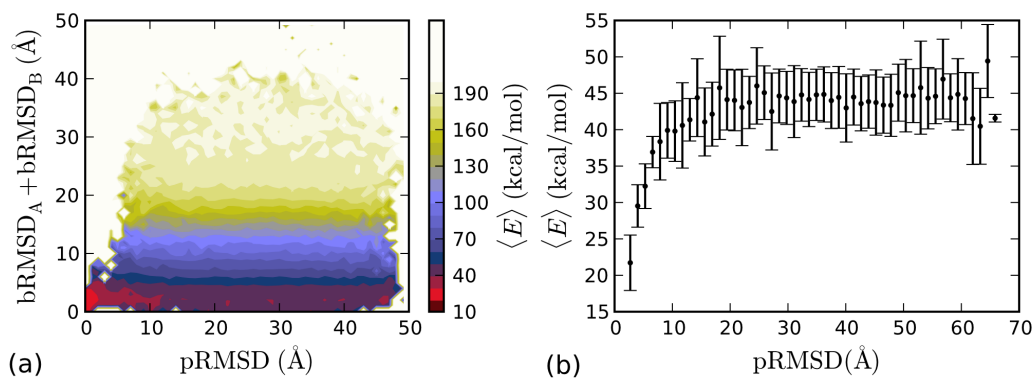}
  \caption{The heterodimeric AB zipper. (a) Mean energy as a function
    of p{RMSD} over both chains and the sum of individual
    b{RMSD}s. The direction of the energy gradients implies
    that a system with two folded monomers is energetically favorable
    compared to unfolded monomers. The proper dimeric form is the area
    closest to the origin, and has a lower energy. (b) Mean energy of
    all states in which both chains have b{RMSD}
    $<$\,5\,\AA, shown as a function of the dimer {RMSD}
    measure p{RMSD}.}\label{fig:eff_8}
\end{figure}

The lowest energy state seen in our simulations is a conformation in
which p{RMSD} calculated over backbone atoms of all residues
in both chains is $\sim$\,2.7\,\AA. In this structure, the
b{RMSD} (all residues) of the individual chains A and B to
their counterparts in the PDB structure are $\sim$\,2.5\,\AA~and
$\sim$\,2.4\,\AA, respectively.  Unlike for the other systems
described in this article, the boundary conditions have a non-trivial
role for this dimeric system. A proper discussion of periodicity,
concentration and temperature dependence of this system is beyond the
scope of this article. In Figure~\ref{fig:eff_8}a, we show the energy
landscape, i.e., the mean energy as a function of two order parameters
for this system. The X-axis shows the measure p{RMSD}
described earlier. The Y-axis represents the sum of the backbone
{RMSD} of the individual chains. p{RMSD} can be
very large even if the sum of b{RMSD}s is small: the two
chains can be folded without making the proper inter-chain
contacts. Indeed, the figure shows that the major energy gradients are
along the Y-axis, showing that it is energetically favorable for both
chains to fold to their respective helical states. The correct dimeric
native state is energetically more favorable by $\sim$\,20 kcal/mol
compared to two folded helices without proper inter-chain
contacts. This is seen more clearly in Figure~\ref{fig:eff_8}b, where
we plot the average energy as a function of p{RMSD} for
states with two folded chains. We also simulated the two chains A and
B of the dimer in isolation. Both chains folded to their native
helical conformations.  The melting temperatures estimated based on
helix content for chains A and B are 314\,K and 313\,K,
respectively. As indicated above, for the dimer, thermodynamic
parameters like $\Tm$ cannot be directly estimated from the present
simulations.

\subsection*{Top7-CFr}
Top7-CFr, the C-terminal fragment of the designed 93-residue
$\alpha/\beta$-protein Top7 \cite{top7orig}, is the most complex of
all molecules studied here. It has both $\alpha$-helix and
$\beta$-strand secondary structure elements, and highly non-local
hydrogen bonds between the N- and C-terminal strands. CFr is known to
form extremely stable homodimers, which retain their secondary
structure till very high temperatures like $371$\,K and high
concentrations of denaturants~\cite{top7cfr}.

In \cite{Mohanty:08a,Mohanty:08b}, an earlier version of our model was
used to study the folding of CFr. The simulations pointed to an
unexpected folding mechanism. The N-terminal strand initially folds as
a non-native continuation of the adjoining $\alpha$-helix. After the
other secondary structure elements form and diffuse to an
approximately correct tertiary organization, the non-native extension
of the helix unfolds and frees the N-terminal residues. These residues
then attach to an existing $\beta$-hairpin to complete the
three-stranded $\beta$-sheet of the native structure. Premature
fastening of the chain ends in $\beta$-sheet contacts puts the
molecule in a deep local energy minimum, in which the folding and
proper arrangement of the other secondary structure elements is
hampered by large steric barriers. The above ``caching'' mechanism,
spontaneously emerging in the simulations, accelerates folding by
helping the molecule avoid such local minima.

The folding properties of CFr, including the above mentioned caching
mechanism, are preserved under the current modifications of the
interaction potential. The center of the native free-energy minimum
shifts from b{RMSD} (all residues) of 1.7\,\AA\ as reported in
\cite{Mohanty:08a} to about 2.2\,\AA. This state remains the minimum
energy state, although the new energy function changes the energy
ordering of the other low energy states. The runs made for this study
(see Table~\ref{tab:eff_6}) found 22 independent folding events. The
free-energy landscape observed in the simulations is rather complex
with a plethora of deep local minima sharing one or more secondary
structure elements with the native structure. They differ in the
registry and ordering of strands and the length of the helix. Longer
runs are required for the MC simulations to correctly weight these
different minima.  Temperature dependence of the properties of CFr can
therefore not be reliably obtained from these runs.

We note that the simulations ran on twice as many processors but were
only about one sixth the length of those used for \cite{Mohanty:08a},
in which 15 independent folding events were found. The improved
efficiency is partly due to the changes in the energy function
presented here, and partly due to the optimization of the parallel
tempering described in \cite{Meinke:09a}.

\subsection*{GS-\balpha$_\text{3}$W}

GS-$\alpha_3$W is a designed three-helix-bundle protein with 67
residues~\cite{Johansson:98}, whose structure was characterized by
NMR~\cite{Dai:02}. The stability was estimated to be $4.6$\,kcal/mol
in aqueous solution at 298\,K, based on CD data~\cite{Johansson:98}.

\begin{figure}
  \includegraphics[width=\textwidth]{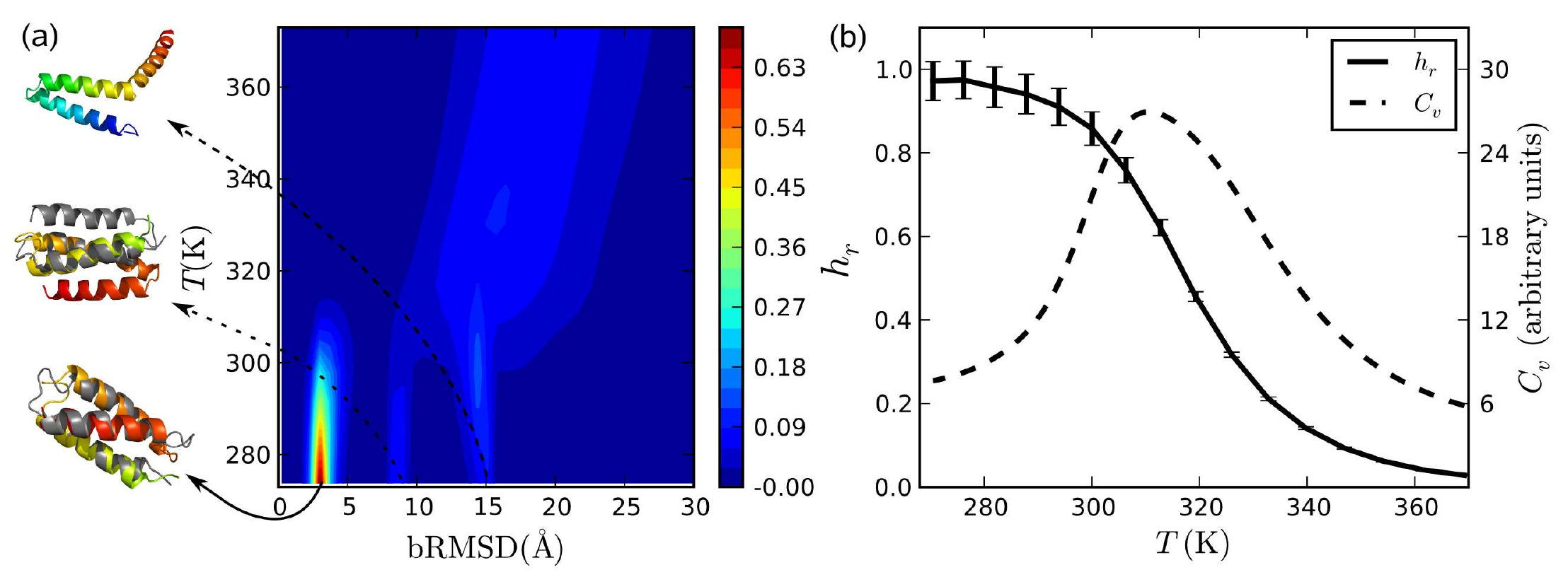}
  \caption{The three-helix-bundle protein GS-$\alpha_3$W. (a)
    Variation of histogram of b{RMSD} with temperature.  At
    high temperatures, there is a broad distribution of
    b{RMSD} with values $>$\,10\,\AA. At lower temperatures
    there are three clearly separated clusters. Representative
    structures from these clusters are also shown (color) aligned with
    the native structure (gray).  (b) Temperature dependence of
    specific heat, $C_v$, and the ratio $h_r$ of the observed helix
    content and the helix content of the native
    structure.}\label{fig:eff_9}
\end{figure}

It turns out that this protein is very easy to fold with our
model. Our results are based on extensive sampling of the conformation
space with $64\times3.5\times10^9$~Monte Carlo updates, resulting in
about 800 independent folding events to the native state. For this
estimate, structures with b{RMSD} (all residues) under
$5$~\AA\ were taken to be in the native minimum (see
Figure~\ref{fig:eff_9} for justification). Two visits to the native
state were considered statistically independent (i) if they occurred
in independent Markov chains, or (ii) if the two visits to the native
state were separated by at least one visit to the highest temperature
in the simulation. For the entire run, we spent about 10 days of
computing time on 64 AMD Opteron processors running at 2.0 GHz.

In Figure~\ref{fig:eff_9}a, we show how the probabilities for
structures with different b{RMSD} vary with temperature in
the simulations. Clearly, the protein makes a transition from a rather
continuous distribution of b{RMSD} at high temperatures to a
distribution dominated by three well separated clusters. Analysis of
the structures at the lower temperatures shows that all three
free-energy minima consist almost exclusively of structures with all
three helices of GS-$\alpha_3$W formed.  The plot of the ratio of the
observed helix content and the helix content of the native state,
shown in Figure~\ref{fig:eff_9}b, further supports this idea. The
average value of this ratio approaches $1$ as the temperature
decreases below 300\,K. The specific heat curve, also shown in
Figure~\ref{fig:eff_9}b, indicates that the formation of these
structures correlates with the steepest change in energy.

The cluster with a center at b{RMSD} $\sim$\,3\,\AA\,
dominates at the lowest temperatures. The structures contributing to
the cluster with $\sim$\,8--9\,\AA\ b{RMSD} superficially
look like well folded three-helix bundles. But as illustrated in the
figure, the arrangement of the helices is topologically distinct from
the native arrangement. The cluster seen at larger b{RMSD}
values is broader and consists of a host of structures in which two of
the helices make a helical hairpin, but the third helix is not bound
to it. The unbound helix could be at either side of the chain.

According to our model therefore, the population at the lowest
temperatures consists of $\sim$\,80\% genuinely native structures,
$\sim$\,10\% three-helix bundles with wrong topology, and $\sim$\,10\%
other structures with as much helix content as the native state. In
order to experimentally determine the true folded population of the
protein, the experimental probe must be able to distinguish the native
fold from the other helix rich structures described here.

\section*{Discussion}

The model presented here is intrinsically fast compared to many other
all-atom models, because all interactions are short range.  By
exploiting this property and using efficient MC
techniques, it is possible to achieve a high sampling efficiency. We
could, for example, generate more than 800 independent folding events
for the 67-residue GS-$\alpha_3$W.  The speed of the simulations thus
permits statistically accurate studies of the global free-energy
landscape of peptides and small proteins.

In developing this potential, a set of 17 peptides with 10--37
residues was studied. The peptides were added to this set one at a
time. To fold a new sequence sometimes required fine-tuning of the
potential, sometimes not.  A change was accepted only after testing
the new potential on all previous sequences in the set.  In its final
form, the model folds all 17 sequences to structures similar to their
experimental structures, for one and the same choice of potential
parameters.

Also important is the stability of the peptides. A small polypeptide
chain is unlikely to be a clear two-state folder, and therefore its
apparent folded population will generally depend on the observable
studied. For $\beta$-sheet peptides, we used the hydrophobicity energy
$\Ehp$ and the hydrogen bond-based nativeness measure $\qhb$ to
monitor the melting behavior. The extracted $\Tm$ values indeed showed
a clear probe dependence; the $\Ehp$-based value was always larger
than that based on $\qhb$. For the $\beta$-hairpins studied, we found
a good overall agreement between our $\Ehp$-based results and
experimental data. For the three-stranded $\beta$-sheets, instead, the
$\qhb$ results agreed best with experimental data. The reason for this
difference is unclear. One contributing factor could be that
interactions between aromatic residues play a more important role for
the $\beta$-hairpins studied here than for the three-stranded
$\beta$-sheets. These interactions may influence spectroscopic signals
and are part of $\Ehp$. Probe-dependent $\Tm$ values have also been
obtained experimentally, for example, for trpzip2~\cite{Yang:04}.

The probe dependence makes the comparison with experimental data less
straightforward. Nevertheless, the results presented clearly show that
the model captures many experimentally observed stability
differences. In particular, among related peptides, the calculated
order of increasing thermal stability generally agrees with the
experimental order, independent of which of our observables we use.

It is encouraging that the model is able to fold these 17
sequences. However, there is no existing model that will fold all
peptides, and our model is no exception. Two sequences that we
unsuccessfully tried to fold are the $\beta$-hairpins trpzip4 and
\Ubi, both with 16 residues.  Trpzip4 is a triple mutant of
{GB1}p with four tryptophans~\cite{Cochran:01}.  For
trpzip4, our minimum energy state actually corresponded to the
NMR-derived native state~\cite{Cochran:01}, but the population of
this state remained low at the lowest temperature studied
($\sim$\,14\,\% at 279\,K, as opposed to an estimated $\Tm$ of 343\,K
in experiments~\cite{Cochran:01}).  \Ubi\ is derived from the
N-terminal $\beta$-hairpin of ubiquitin~\cite{Jourdan:00}. It has a
shortened turn and has been found to form a $\beta$-hairpin with
non-native registry~\cite{Jourdan:00}. In our simulations, this state
was only weakly populated ($\sim$\,8\% at 279\,K, as opposed to an
estimated $\sim$\,80\% at 288\,K~\cite{Jourdan:00}).  Instead, the
main free-energy minima corresponded to the two $\beta$-hairpin states
with the registry of native ubiquitin, one with native hydrogen bonds
and the other with the complementary set of hydrogen bonds.

Our calibration of the potential relies on experimental data with
non-negligible uncertainties, on a limited number of peptides.  It is
not evident that this potential will be useful for larger polypeptide
chains. Therefore, as a proof-of-principle test, we also studied three
larger systems, with very good results. Our simulations showed that,
without having to adjust any parameter, the model folds these
sequences to structures consistent with experimental data. Having
verified this, it would be interesting to use the model to investigate
the mechanisms by which these systems self-assemble, but such an
analysis is beyond the scope of this article. The main purpose of our
present study of these systems was to demonstrate the viability of our
calibration approach.

The potential can be further constrained by confronting it with more
accurate experimental data and data on new sequences. The challenge in
this process is to ensure backward compatibility --- new constraints
should be met without sacrificing properties already achieved.

\section*{Conclusion}

We have described and tested an implicit solvent all-atom model for
protein simulations. The model is computationally fast and yet able to
capture structural and thermodynamic properties of a diverse set of
sequences.  Its computational efficiency greatly facilitates the study
of folding and aggregation problems that require exploration of the
full free-energy landscape. A program package, called
PROFASI~\cite{profasi}, for single- and multi-chain simulations with
this model is freely available to academic users.

\subsection*{Acknowledgements}

We thank Stefan Wallin for suggestions on the manu\-script. This work
was in part supported by the Swedish Research Council. The simulations
of the larger systems were performed at the John von Neumann Institute
for Computing (NIC), Research Center J\"ulich, Germany.

\bibliographystyle{physmath_art} 
  \bibliography{eff} 

\end{document}